\documentclass{article}
\usepackage{fleqn}
\usepackage{espcrc2}
\usepackage{epsf}
\usepackage{epsfig}
\usepackage{rotating}

\title{
Local bosonic versus HMC --- 
a CPU cost comparison
}

\author{
S. Elser 
\thanks{\mbox{supported by DFG research grant No. WO 389/3-2;} 
        \mbox{$\,$ email: elser@linde.physik.hu-berlin.de;}
        \mbox{$\,$ transparencies:
        http://linde.physik.hu-berlin.de/elser}
}
and B. Bunk 
\thanks{email: bunk@linde.physik.hu-berlin.de} \\
{Institut f\"ur Physik, Humboldt--Universit\"at, Invalidenstr.110,
         10115 Berlin, Germany 
        }
       }
       
\begin{document}


\begin{abstract}
We compare 
the CPU cost of HMC
to various implementations of the
Hermitean variant of L\"uscher's local bosonic algorithm (LBA)
for 2D massive QED with two flavours of Wilson
fermions.
We carefully scan a 3-dimensional parameter subspace
and find flat behaviour around the optimum.
The gain factor 
of
 the LBA,
as compared to HMC,
is
slightly smaller
for the Re-weighting method
than for the Metropolis variants
and estimated
to
about 3.3 for the plaquette
and 1.9 for the the meson correlator.
\end{abstract}

\maketitle


\section{Schwinger model}

For our tests we select
as a low-cost laboratory
the
massive 2-flavour Schwinger model 
(2D QED)
\cite{sch62,het95}.
The
lattice model is given
by the
Wilson plaquette and fermionic action.
For details we refer to last year's proceedings \cite{els96}.
We work with the
Hermitean
fermion matrix
$
Q = 
c\gamma_5 M
$ scaled so that its 
eigenvalues are in $[-1,1]$.


\section{Hybrid Monte Carlo}

To set the scale,
we simulate 
using 
a HMC code
working also with the Hermitean matrix $Q$.
The implementation
includes optimization features like
trajectory length set by
{ $n \cdot \Delta\tau = 1$}
and 
{ acceptance $\approx 70 \%$},
and
re-use of the CG solution in the trajectory
via $\stackrel{\dots}{x}=0$ (gain $\approx$ 20\%).


\section{Hermitean local bosonic algorithm}

Alternatively to HMC,
M. L\"uscher proposed a local bosonic formulation \cite{lue94}.
The one main variant we consider
uses the
fact that we are dealing with a Hermitean fermion matrix
thus
allowing
to
exactly
rewrite
the effective distribution 
\begin{eqnarray}
P_{\rm eff} 
&\propto& 
{\rm det} Q^2 \, e^{-S_g(U)} 
\nonumber \\
&\propto& 
{\rm det}[ 1-R] \;\;
{\rm det}[P_n(Q^2)^{-1}] \;\;
e^{-S_g(U)} 
\nonumber
\end{eqnarray}
with
$P_n(s)$ 
a polynomial of even degree $n$
approximating
${1 \over s}$  for {\em real} $s \in [\epsilon,1]$ 
such that the correction factor
${\rm det}(1-R)=
{\rm det}[ Q^2
P_n(Q^2)] 
 \approx 1$.  
Its roots $z_k$ ($k=1 \ldots n$) come in complex conjugate pairs
and determine $\sqrt{z_k}=\mu_k + i\nu_k$ ($\nu_k>0$).
This leads to a
{ totally bosonic} representation
\begin{eqnarray}
P_{\rm eff} 
\propto \hspace{6.4cm}
\nonumber
\\
{\rm det}[ 1-R]
e^{-S_g(U)} 
\int {\cal D}\phi \, e^{-\sum_k \phi_k^\dagger [ (Q-\mu_k)^2 
+\nu_k^2] \phi_k} 
\nonumber
\end{eqnarray}
with $n$ complex bosonic Dirac fields $\phi_k$.
We chose as approximation polynomials
$P_{n}(s)$ the Chebyshev polynomials proposed by 
Bunk et al.\cite{bun95}. The convergence of
$P_{n}(s) \rightarrow 1/s$ as $n \rightarrow \infty$ is
exponential and uniform for $s \in [\epsilon,1]$.
One update of the bosonic system consists of
a trajectory of heat bath 
and over-relaxation steps
for $U$ \cite{bes79} and $\phi$ fields,
possibly followed by a Metropolis appectance correction
for $\det[1-R]$.
In total, this introduces 
the 
parameters
$n$, $\epsilon$
and the number of
reflections
per heat bath 
into the algorithm.


\section{Exact algorithm}

The correction factor
can 
be treated exactly with
a
Metropolis
accept/reject step
or Re-weighting \cite{lue96}
using a 
stochastic method
as demonstrated 
for the non-Hermitean variant
\cite{for96}.

The Reweighting method computes a noisy estimate for 
$\det[1-R]$ and includes it in the observables 
.

The Metropolis correction step uses an
 acceptance probability
$P^A(\chi)$ dependent on a Gaussian noise $\chi$
which
satisfies detailed balance when averaged over the noise.
This can be achieved formally
by
\begin{eqnarray}
P^A_{U \phi \to U' \phi'}
&=&
{\rm min} \bigl( 1,e^{- \chi^\dagger B^\dagger [1-R']^{-1} B \chi +
\chi^\dagger \chi } \Bigr)
\nonumber
\end{eqnarray} 
with $B$ given by
$
B B^\dagger = Q^2 P_n(Q^2)
$.
In the Hermitean case this is non-trivial 
to solve.
The task becomes trivial, if we recall
the factorized form
\begin{eqnarray}
Q^2 P_n(Q^2)
 = Q^\dagger Q \,\, N_{\rm norm}
\prod_k^{n/2} (Q^2-z_k) (Q^2-\bar z_k)
\nonumber
\end{eqnarray}
taking one factor 
from each c.c. pair.

We remark that the Metropolis scheme
includes a further
optimization possibility which we call
Metropolis with adapted precision.
We retain a valid algorithm if
the inversion necessary for the Metropolis decision
is first executed with very small precision (in our case
$\delta=10^{-2}$) and repeated with standard inverter 
precision ($\delta=10^{-6}$)
only if the decision would else be unclear.
As
the CG can be restarted
from the intermediate solution,
this procedure
could result in less work on the average.


\section{Numerical instabilities}

Evaluating a high order polynomial
for a matrix faces the problem of
loss of precision.
In the non-Hermitean variant 
or Re-weighting case
we are able to avoid this by using the
Chebyshev recursion formula for $R$.
Unfortunately
the
Hermitean variant with
Metropolis correction 
or the Polynomial Hybrid Monte-Carlo algorithm \cite{jan97}
rely on the evaluation of partial products of root factors.

In a forthcoming publication \cite{els97},
we
compare various proposals
for reordering the roots
to minimize 
numerical instabilities.
In this work,
we use a fairly stable version,
the so-called
Bit\-rever\-sal scheme, whenever no recursion
is possible.

Recently, one of us
proposed
a different solution \cite{bun97}.
We rewrite the 
Metropolis acceptance
to
\begin{eqnarray}
P^A_{U \phi \to U' \phi'}
&=&
{\rm min} 
\left( 1,
e^{\eta^\dagger (R-R') \eta } 
\right)
,
\nonumber
\end{eqnarray}
with $\eta$ given by
$\chi = [Q^2 P_n(Q^2)]^{1 \over 2} \eta$.
At this point we suggest 
solving for the inverse of the square root of
$Q^2 P_n(Q^2)$ directly.
This can be accomplished
with an expansion
in Gegenbauer polynomials
converging exponentially 
with the same rate as CG.


\section{Observables -- $\kappa$ value }

We simulate on
8x20 lattices with a conservative
$\beta=3.0$ 
and a 
run length of usually  $>1000 \tau$,
calculating
the plaquette and 
correlations of local
meson operators
$\langle \bar\Psi \gamma^5 \tau \Psi \rangle$ ($\pi$), 
$\langle \bar\Psi \tau \Psi \rangle$ ($a_0$)
,
$\langle \bar\Psi \gamma^5 \Psi \rangle$ ($\eta$).
We expect finite size effects
to appear in deviations
from the approximately linear behaviour
of the pion mass with $\kappa$.
As shown in Fig. \ref{mass},
finite size
effects are small for $\kappa \le 0.24$,
resulting in a
pion mass 
$m_\pi$ = 0.629
and 
a physical ratio { ${m_\pi \over m_\eta}= 0.807$}.

\begin{figure}
\vspace{-0.5cm}
\label{mass}
\caption{}
\epsfig{file=notes_060797_search.data8_4_ps, width=8cm}
\vspace{-1cm}
\end{figure}


\section{Cost comparison}

To summarize, we repeat the
schemes included in our investigation.
Besides
 Hybrid MC to set the scale,
we compare
 LBA with { Re-weighting},
 LBA with { Metropolis} ,
 LBA with { Metropolis using adapted precision},
and
 LBA with { Gegenbauer inverter}
as described above.

We point out
that
Re-weighting and Metropolis algorithms
result in
2 different ensembles.
Therefore
it is no longer possible
to compare CPU cost by
\begin{eqnarray}
C
=
{2 \tau_{\rm int} \over N_{\rm meas}} \cdot N_{\rm all \; Q \; ops}
\nonumber
\end{eqnarray}
but
one has to use a measure based on the relative error
\begin{eqnarray}
C_{\rm eff}
=
N_{\rm all \; Q \; ops} \cdot {\sigma_{\rm tot}^2 (A) \over <A>^2}.
\nonumber
\end{eqnarray}
In these formulae $N_{\rm meas}$ 
signifies the number of measurements,
$ N_{\rm all \; Q \; ops}$
the total number
of matrix multiplications for the whole run.

To illustrate our 
search for optimal parameters,
we depict in Fig. \ref{cont}
the 
CPU cost
in the 
$n-\epsilon$ plane
(number of reflections optimized)
for one algorithm,
namely 
plain Metropolis.
The figure clearly shows that we obtain 
a flat
optimum.

\begin{figure}
\vspace{-0.5cm}
\caption{}
\label{cont}
\hspace{-0.8cm}
\epsfig{file=notes_060697_8x20_CPU.data120_3_ps, width=8cm}

\vspace{-1.5cm}
\end{figure}

We further compare only the optimal
parameter sets and their CPU cost in table \ref{cost}.

\begin{table}
\caption{CPU cost minina - Plaquette}
\label{cost}

\begin{tabular}{|c|c|c|c|c|c|}
\hline\hline
algorithm & $n$ & $\epsilon$ & refl. & cost & gain 
\\ \hline
HMC & & & & 6.1 &
\\ \hline\hline
Metro 1 & 18 & 0.02 & 2 & 1.9 & 3.3
\\ \hline
Metro 2 & 18 & 0.02 & 3 & 2.1 & 2.9
\\ \hline
Gegenbauer & 18 & 0.01 & 2 & 2.0 & 3.0
\\ \hline
Re-weighting & 24 & 0.02 & 4 & 2.6 & 2.4
\\

\hline\hline
\end{tabular}
 
\vspace{0.3cm}
Pion Propagator
\vspace{0.1cm}

\begin{tabular}{|c|c|c|c|c|c|}
\hline\hline
algorithm & $n$ & $\epsilon$ & refl. & cost & gain 
\\ \hline
HMC & & & & 574 &
\\ \hline\hline
Metro 1 & 24 & 0.01 & 2 & 501 & 1.1
\\ \hline
Metro 2 & 18 & 0.04 & 1 & 320 & 1.8
\\ \hline
Gegenbauer & 12 & 0.04 & 1 & 304 & 1.9
\\ \hline
Re-weighting & 18 & 0.02 & 4 & 414 & 1.4
\\

\hline\hline
\end{tabular}
\vspace{-0.3cm}
\end{table}

\section{Conclusions}

We think that this study
gives further clear evidence
that the
LBA is competitive to HMC.
The tuning of the LBA is fairly easy.
The CPU cost can be
lower than for HMC,
but not by a large factor
with present techniques.
We want to stress that
the gain factors for the plaquette
and pion propagator 
differ 
with estimates from plaquette-like
observables too optimistic.

Technically,
the number of reflections
per heat-bath update
is found to be an important optimization tool.
We also demonstrate
that using a
Noisy Metropolis scheme
to make the LBA exact is
possible
for the Hermitean case as well.
The Gegenbauer inverter,
which
avoids instabilities
in the evaluation of the polynomial,
is shown to be 
competitive to CG
in a first real simulation.


\section{Acknowledgements}

We thank U. Wolff, C. Lang,
K. Jansen and R. Frezzotti for discussions.


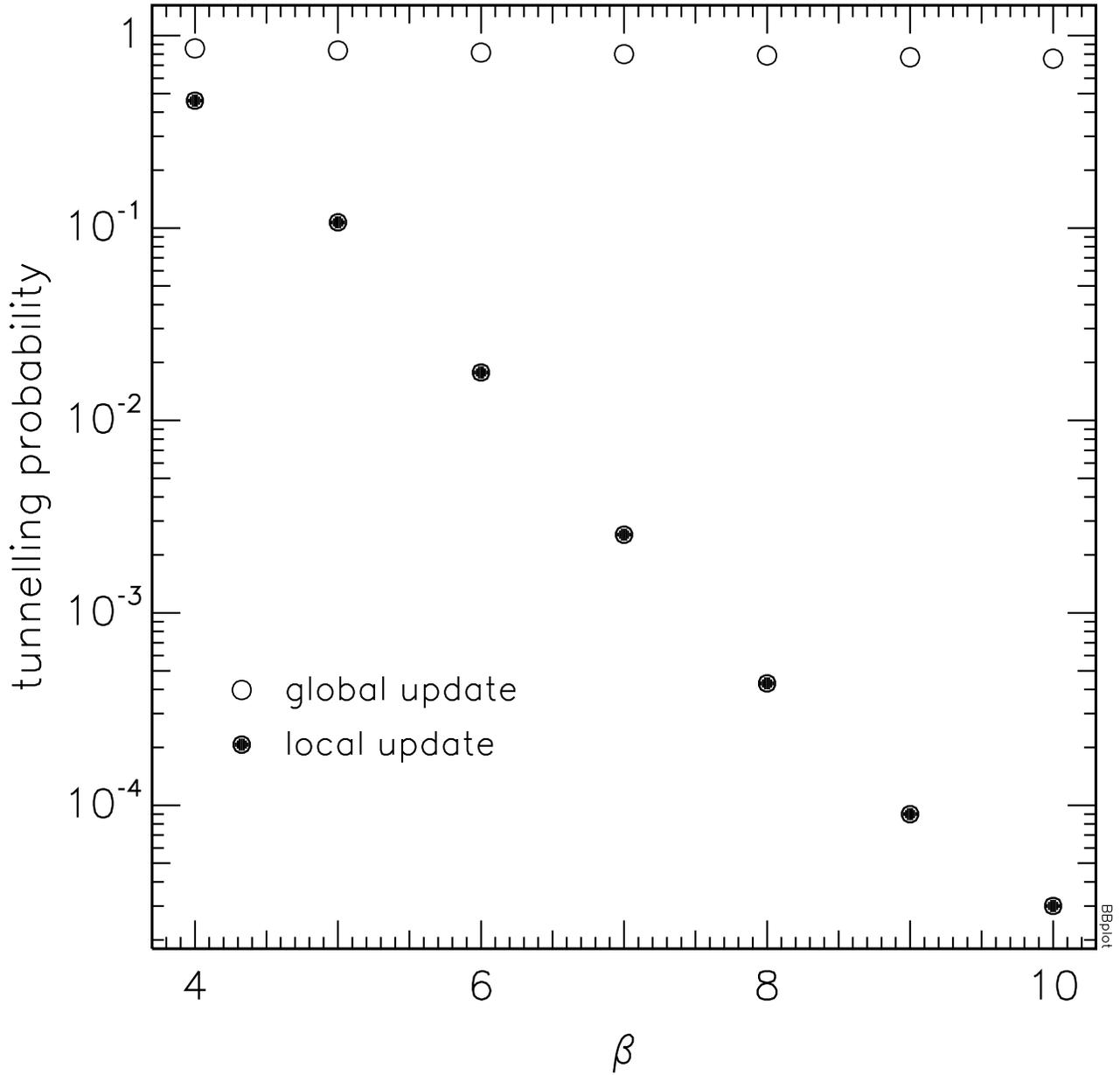
\end{document}